\documentclass[sigconf,10pt]{acmart}

\usepackage[linesnumbered,ruled,vlined]{algorithm2e}
\usepackage{balance}
\usepackage{booktabs}
\usepackage{caption}
\usepackage{enumitem}
\usepackage{flushend}
\usepackage{float}
\usepackage{graphicx}
\usepackage{hyperref}
\usepackage{makecell}
\usepackage{multirow}
\usepackage{mathtools}       
\usepackage{pifont}
\usepackage{subcaption}
\usepackage{tabularx}
\usepackage{tabularray}
\usepackage{xcolor}
\usepackage[most]{tcolorbox}
\usepackage{placeins}
\definecolor{darkgreen}{RGB}{0,128,0} 
\definecolor{darkred}{RGB}{200,0,0}  

\renewcommand\footnotetextcopyrightpermission[1]{} 
\newcommand{\cmark}{\textcolor{darkgreen}{\ding{51}}} 
\newcommand{\xmark}{\textcolor{darkred}{\ding{55}}}   

\newcommand{\name}{Synera}

\begin{document}

\title{\name: Synergistic LLM Serving across Device and Cloud at Scale}



\author{Genglin Wang\textsuperscript{†,*},
Liekang Zeng\textsuperscript{†,*},
Bufang Yang\textsuperscript{†},
Kaiwei Liu\textsuperscript{†}, \\
Guoliang Xing\textsuperscript{†},
Chumin Sun\textsuperscript{‡},
Li Zhou\textsuperscript{‡},
Jie Sun\textsuperscript{‡},
Zhenyu Yan\textsuperscript{†}}

\affiliation{
\textsuperscript{†}The Chinese University of Hong Kong \city{Hong Kong SAR}
\country{China}\\
\textsuperscript{‡}Theory Lab, Central Research Institute, 2012 Labs, Huawei Technologies Co. Ltd.
\country{China} \\\
\textsuperscript{*}Equal contribution
}










\renewcommand{\shortauthors}{xxx et al.}

\begin{abstract}
Large Language Models (LLMs) are becoming key components in various mobile operating systems, driving smart applications like interactive chatbots and personal assistants.
While bringing enhanced intelligence to mobile ends, their deployment suffers from a set of performance challenges, especially the generation quality degradation and prolonged latency.
Prior works have mainly relied on solutions of cloud offloading or on-device Small Language Models (SLMs).
However, the former is usually limited by the communication bottleneck, and the latter sacrifices generation quality due to resource constraints.
To mitigate these limitations, this paper proposes Synera, a device–cloud synergistic LLM serving system that applies an efficient SLM-LLM synergistic mechanism.
Through empirical studies on LLM's unique computing characteristics, Synera identifies a set of underexplored optimization opportunities in device-cloud synergistic LLM inference, including offloading decisions, pipeline stalls, and batching bottlenecks.
To translate them into enhanced performance, Synera introduces tailored designs of communication-efficient selective offloading, stall-free parallel inference, and scalable cloud batching.
Extensive evaluations with real-world testbeds show that Synera enables 1.20–5.47$\times$ better generation quality against competitive baselines with on-par latency performance.
Compared with existing cloud serving, Synera achieves 8.2-16.5\% lower cloud serving cost on various benchmarks.

\end{abstract}

\maketitle

\section{Introduction}

Large language models (LLMs) are becoming a cornerstone for a wide range of mobile and edge LLM services, including mobile agent~\cite{small_language_model_SLM_slm}, mobile health~\cite{bufang_health}, and autonomous driving~\cite{zheng2025driveagent}.
However, the computational demands and massive parameters hinder their deployment for resource-constrained mobile and edge devices. In practice, such devices can only sustain models up to the small-scale (<~10B parameters), far from the hundred-billion–parameter regime that delivers state-of-the-art performance.
Current approaches to address this mismatch fall into two categories.
The first is cloud offloading, where devices offload requests to a cloud server and leverage abundant resources for high-quality generation, but require stable connectivity and incur substantial operational costs.
The second is the adoption of on-device Small Language Models (SLMs) through techniques such as compression, which improve deployability and reduce connectivity concerns, but are intrinsically constrained by the device and thus cannot approach the capability of large-scale LLMs. Our measurement shows that relying solely on on-device SLMs may cause up to 58\% accuracy degradation relative to cloud-based LLMs.

\begin{figure}[!t]
    \centering
    \includegraphics[width=0.95\linewidth]{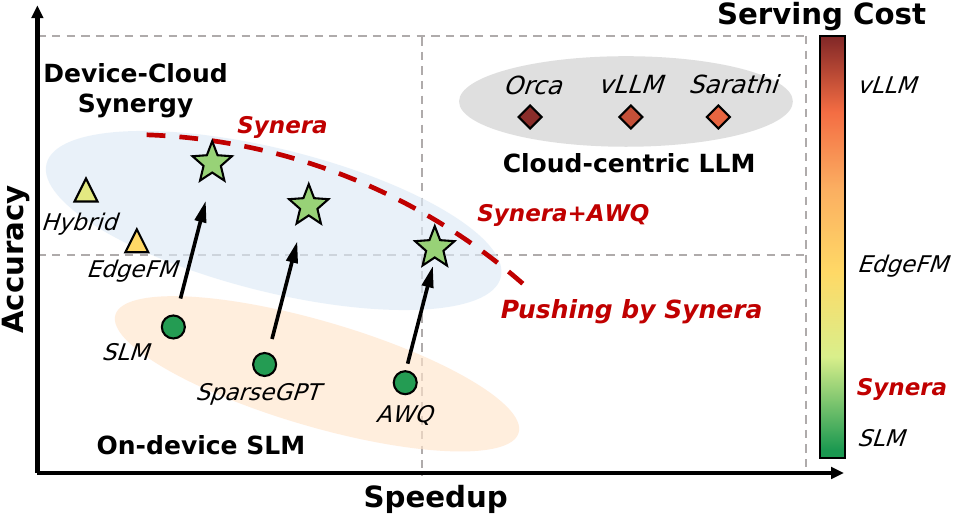}
    \caption{Current LLM serving involves on-device SLM, cloud-centric LLM, and device-cloud synergy. \name{} notably advances the accuracy-speedup frontier with significantly lower serving costs.}
    
    \label{fig:intro}
\end{figure}

To overcome the intrinsic ceiling of devices, recent work has explored device-cloud synergistic serving, which jointly schedules computation across device and cloud to harness their complementary strengths. This paradigm encompasses a range of strategies, including data compression to reduce transmission latency~\cite{deepcod}, multi-stage processing that splits lightweight and heavyweight tasks~\cite{hung2018videoedge}, neural network (NN) partition that splits NN layers between device and cloud for flexible workload balancing~\cite{SPINN}, and small-large model synergy that selectively offloads challenging inputs from a on-device small model to a powerful cloud-based model~\cite{edgefm}. Together, these approaches demonstrate the potential of synergy in improving efficiency and accuracy beyond what either a standalone device or cloud serving can achieve.

\textbf{Challenges}. While these explorations operate well on traditional AI models like convolutional visual models \cite{SPINN,zeng2020coedge,zeng2019boomerang}, directly applying them to LLM may fall short in serving performance.
First, the next-token prediction of LLM is fundamentally distinct from bulk data processing, such as image and video processing, where data compression and multi-stage methods are predominantly applied. Consequently, such two approaches are generally ineffective for LLM.
Second, auto-regressive generation requires frequent per-token model forward passes. If the NN partition is used, this can expose the LLM service to frequent communication overhead, making it difficult to guarantee latency constraints.
Third, generative LLMs decode a sequence of tokens, whereas traditional discriminative models output a single class label for each input sample. This difference in output granularity creates a challenge for small-large model synergy, which typically relies on per-sample offloading. As a result, existing synergistic serving falls short in delivering accurate, fast, and scalable generative LLM inference.

\textbf{Observation.} We observe that SLM-LLM synergy, where an on-device SLM collaborates with a cloud-based LLM, offers a promising pathway for efficient and scalable LLM serving. More importantly, our key insight is to enable \textit{token-level synergy}: selectively offloading only those token chunks beyond the SLM’s capacity, i.e., the quality-critical (``hard'') token chunks, to the cloud LLM for verification and refinement. This approach, inspired by the ``draft \& verify'' paradigm in speculative decoding~\cite{leviathan2023fast_google_speculative, chen2023accelerating_deepmind_speculative}, effectively leverages the strengths of both models while minimizing unnecessary communication. It offers clear benefits in accuracy, latency, and cost, and its flexible design can be integrated with established privacy-preserving techniques to mitigate privacy concerns~\cite{yuan2025scx,dong2023puma}.
However, it introduces new challenges, such as precisely identifying quality-critical tokens and maximizing the efficiency of both device and cloud resources.

\textbf{\textsc{\name{}}}. In this paper, we propose \name{}, a distributed device-cloud syneretic system that delivers high-quality, low-latency, and cost-efficient LLM serving at scale.
\name{} applies a hybrid SLM-LLM serving mechanism via token-level collaboration, where the mobile devices perform SLM for in-situ resilient model inference with the cloud hosts LLM for generation enhancement. 
Specifically, \name{} introduces several key technical innovations: 
(1) \textbf{Selective token-level offloading} that combines SLM's confidence and attention-based importance scores for differentiated token transmission. By using the offloading budget as a knob, this selection method enables tunable quality–efficiency trade-offs.
(2) \textbf{Progressive early exit inference}, integrating layer-wise and sequence-wise early exit strategies on the device. This reduces local overhead and also provides signals for offloading decisions.
(3) \textbf{Stall-free parallel inference} that overlaps computation and communication in the collaborative pipeline across device and cloud, and effectively unleashes on-device pipeline efficiency. 
(4) \textbf{Efficient cloud scheduling}.
To support a multi-user environment at scale, we further introduce an efficient scheduler for throughput-optimized continuous batching~\cite{yu2022orca}.
This module reinterprets non-uniform requests into uniform partial prefill, simultaneously improving cloud resource efficiency and reducing operational cost.

Assembling these key designs, \name{} can deliver flexible device-cloud synergy of LLMs with significant response quality improvement, while allowing both communication and serving cost efficiency. 
Fig. \ref{fig:intro} and Table~\ref{tab:comparison with related work} show the positioning and superiority of \name{} over state-of-the-art related systems.
We summarize our key contributions as follows:

$\bullet$ We propose token-level synergy, which selectively offloads only quality-critical tokens, thereby achieving an overall improvement in accuracy, latency, and serving cost.

$\bullet$ We design \name{}, a synergistic and scalable LLM serving system across device and cloud that leverages selective token-level offloading, progressive early exit inference, stall-free parallel inference, and efficient cloud scheduler for optimal on-device and cloud efficiency.

$\bullet$ We extensively evaluate \name{} on diverse real-world tasks across edge and mobile testbeds, demonstrating up to 5.47$\times$ improvement in generation quality and a reduction in cloud serving cost to 8.2–16.5\%. \name{} outperforms competitive baselines in end-to-end accuracy, speed, and cost.

\section{Background}

\begin{table}[!t]
\centering
\resizebox{\linewidth}{!}{%
\begin{tabular}{l|cccc}
\toprule
\textbf{Method}          & \makecell{\textbf{LLM-} \\ \textbf{compatible}} & \makecell{\textbf{Serving Cost} \\ \textbf{Reduction}} & \makecell{\textbf{Comm.} \\ \textbf{Efficiency}} & \makecell{\textbf{Multi-user} \\ \textbf{Scalability}} \\
\midrule
DeepCOD~\cite{deepcod}          & \xmark & Low & High & \xmark \\
Edgent~\cite{li2019edge}        & \xmark & Low & Medium & \xmark \\
SPINN~\cite{SPINN}              & \xmark & Medium & Medium & \xmark \\
EdgeFM~\cite{edgefm}            & \xmark & Medium & Medium & \xmark \\
CollabTrans~\cite{chen2025collabtrans}       & \xmark & Medium & Medium & \cmark \\
Hybrid~\cite{native_hybrid_renju_zixuhao}         & \cmark & Medium & Low & \xmark \\
\midrule
\textbf{\textsc{\name{}}}      & \cmark & High & High & \cmark \\
\bottomrule
\end{tabular}
}
\caption{Comparison of \name~and related work for device-cloud synergistic NN and LLM inference.}\label{tab:comparison with related work}
\end{table}

\subsection{Device-cloud Synergy}\label{background: model partition}

Device–cloud synergy is a mainstream paradigm for on‑device model serving. To harness the complementary strengths of device and cloud, prior work explored techniques spanning computation and communication. Early studies investigated data compression for reducing communication overhead~\cite{deepcod} and multi‑stage processing for pipelining functional stages (e.g., of video analytics) across device and cloud~\cite{hung2018videoedge}.
These methods are generally designed for synergy acceleration without tailored optimization for AI applications.

To enable efficient device-cloud synergy for AI models, recent work has mainly advanced along two directions: NN partition and small-large model synergy. Device-cloud NN partition~\cite{gaowei_AgileNN,SPINN,chen2025collabtrans,li2019edge,huang2023re} splits a neural network with the initial few layers deployed on the device and the remaining on the cloud. Representative work such as SPINN~\cite{SPINN} co-optimizes the early exit policy and splitting point to maximize efficiency in both speed and accuracy. In contrast, small–large model synergy~\cite{edgefm,li2021appealnet,native_hybrid_renju_zixuhao} leverages a compact device-side model in tandem with a powerful cloud model for improved efficiency. For example, EdgeFM~\cite{edgefm} performs open-set recognition by retaining high-confidence predictions locally while offloading uncertain cases to the cloud.

\subsection{Challenges of LLM Serving}\label{background: challenges llm}

The above optimization genres have demonstrated their effectiveness in traditional NN inference. 
When deploying LLM in edge scenarios, however, most of them fall short due to a set of LLM-specific challenges, including computation intensity, computation dependency, and computation complexity.

\textbf{Computation intensity.}
The substantial memory required by LLM makes traditional device-cloud serving, particularly for NN partition, impractical on resource-constrained devices. For instance, Qwen-3-235B~\cite{qwen3technicalreport} comprises 94 layers. Each layer contains approximately 2.5B parameters. Consequently, in the context of NN partition, even partitioning a single layer remains prohibitive on many devices. Moreover, even if we can deploy a few layers on the device via quantization, placing only a handful out of hundreds of layers on the device yields negligible benefits and often leads to poor efficiency. Hence, the extreme computational intensity of LLMs fundamentally limits device-cloud collaborative serving.

\textbf{Computation dependency.}
The computation dependency of the auto-regressive LLM poses a significant challenge. Generative LLM forwards hidden states through layers, samples the next token, appends this token to the prompt, and iteratively repeats this process. In this workflow, each output token requires a complete forward pass through transformer layers. For synergistic serving, like NN partition, this strict sequential dependency results in huge communication overhead, as layer hidden states must be transferred between partitioned model segments for every split point at each step. Since hidden states are typically high-dimensional and large in size~\cite{chen2025collabtrans}, transmitting them repeatedly amplifies the overhead, making efficient partitioning particularly impractical.

\textbf{Computation complexity.}
The complexity of synergistic LLM serving comes from the auto-regressive generation, which hinders both NN partition and small-large model synergy. In NN partition, auto-regression demands synchronized execution for every token across the device and cloud, creating high scheduling overhead. Although cross-device LLM parallelism has been studied~\cite{yeshengyuan_jupiter,ye2024asteroid}, its practicality in the real world remains uncertain and limited. For traditional small-large model synergy that conducts input-level offloading, the main issue is output granularity: LLMs generate sequences token by token, while vision models output a single class label. This mismatch makes per-sample offloading ineffective.

Meanwhile, to mitigate the complexity of auto-regression, recent cloud-based work has introduced continuous batching~\cite{yu2022orca}, which schedules the prefill and decode requests during LLM generation at each generation step. However, device-cloud synergistic serving has largely developed in isolation from these state-of-the-art cloud-based systems and therefore cannot fully exploit the cloud-side efficiency.

\subsection{Design Goals}\label{background: goal}
Regarding the above challenges, our system should meet the following design goals for synergistic LLM serving:

\begin{figure}[!t]
    \centering
    \includegraphics[width=0.95\linewidth]{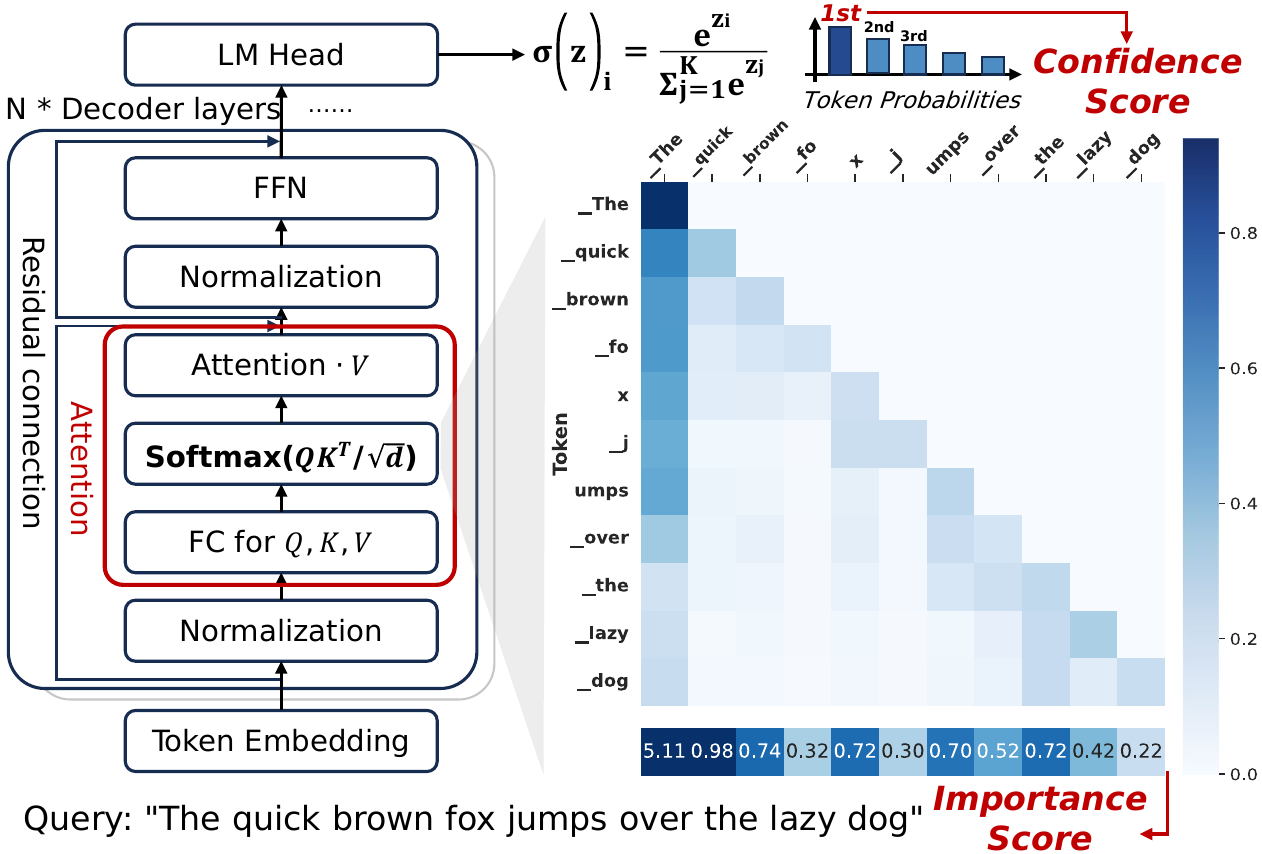}
    \caption{
        An illustration of LLM generation, such as the attention module, and confidence and importance score.
    }
    \label{fig:workflow}
\end{figure}

(1) \textbf{Flexibility.} 
The system should support diverse device–cloud synergy configurations and integrate seamlessly with existing cloud-based LLM serving frameworks (e.g., vLLM~\cite{vllm}). This enables leveraging cloud efficiency while minimizing deployment complexity.

(2) \textbf{Serving cost reduction.}
Computation should primarily remain on devices, with cloud offloading invoked only when locally generated tokens are uncertain. The cloud must further manage serving streams in a cost-efficient manner.

(3) \textbf{Communication efficiency.}
Only data or tokens essential for auto-regressive generation should be transferred between the device and cloud, avoiding redundant communication and improving efficiency.

(4) \textbf{Multi-user scalability.} 
In multi-user settings, the cloud component must scale to handle increasing device requests while sustaining high performance and reliability.

\section{EXPLORING TOKEN-LEVEL SYNERGY}\label{motivation}

We motivate our designs through empirical observations using Llama-68M~\cite{miao2024specinfer} as the SLM and Llama-7B~\cite{touvron2023llama} as the LLM.
We first explore the token-level synergy in Section~\ref{motivation: opportunities and key insight}. Building upon this core idea, we identify bottlenecks of the device and cloud in Section~\ref{motivation: pipeline stalls} and \ref{motivation:cloud_serving}, respectively. 

\subsection{Token-level Synergy}\label{motivation: opportunities and key insight}

\textbf{Opportunity}.
We identify small-large model synergy as a promising opportunity. To be specific, we let on-device SLM and cloud-based LLM collaborate. 
On the one hand, recent advances have significantly improved the capabilities of SLMs, which makes SLM increasingly attractive. This allows SLM to serve as a robust fallback to ensure the availability of local results, even when the cloud connectivity is lost or in offline scenarios.
On the other hand, utilizing the cloud for assistance enables significant improvement of generation quality, given that SLM lags behind LLM in most cases.

\begin{figure}[t]
    \centering
    \includegraphics[width=1.0\linewidth]{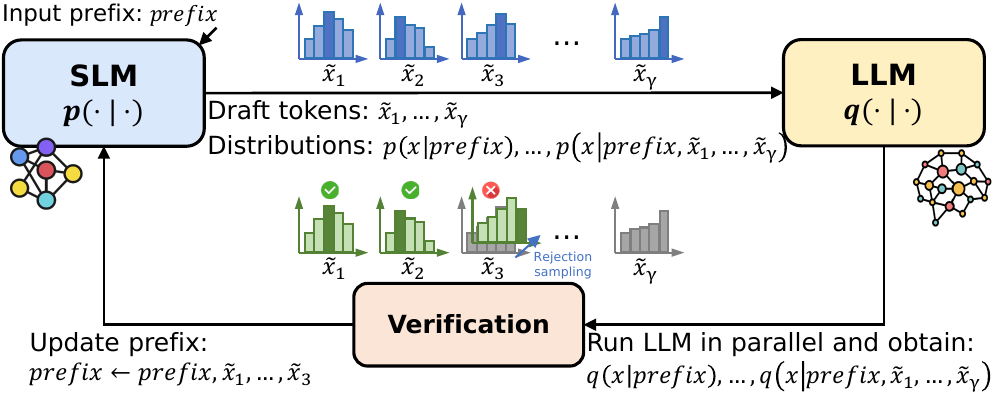}
    \caption{
        An illustration of the ``Draft \& verify'' in speculative decoding.
    }\label{fig:speculative}
\end{figure}

\textbf{Core idea}. We explore \textit{token-level synergy} between on-device SLM and cloud-based LLM. In this workflow, we adopt SLM-centric generation, offloading only ``hard'' tokens to the cloud-based LLM. The LLM acts as a verifier to verify and refine the offloaded outputs to ensure optimal generation quality. This approach is motivated by the increasingly capable SLM that can produce optimal tokens in most cases~\cite{leviathan2023fast_google_speculative,chen2023accelerating_deepmind_speculative}, but may produce bad results at certain critical steps.
This idea is inspired by the “draft \& verify” mechanism (Fig. \ref{fig:speculative}) in speculative decoding~\cite{leviathan2023fast_google_speculative,chen2023accelerating_deepmind_speculative}, where SLM can speculatively draft the LLM prediction and then verify it by the LLM. 

This core idea holds potential in reducing the complexity of LLM synergistic serving, minimizing the communication frequency, and decreasing overall cloud cost by invoking cloud-based LLM only when necessary. However, it also introduces non-trivial challenges. \S~\ref{motivation: confidence and importance} focuses on how to select critical token chunks to offload. \S~\ref{motivation: pipeline stalls} and \S~\ref{motivation:cloud_serving} respectively identify two bottlenecks in the device pipeline and cloud batching, and their corresponding opportunities.

\subsection{Offloading Decision Metrics}\label{motivation: confidence and importance}

\textbf{Challenges.} Discriminative models such as CNNs can estimate prediction confidence via confidence score, i.e., the top-1 softmax probability. Based on this, existing small-large model synergy studies confidence-guided input-level offloading, which selects ``hard'' samples beyond small models' capability to offload~\cite{wang2023tabi,edgefm}. In contrast, generative LLMs decode sequences of tokens, and this finer output granularity is misaligned with existing small–large model synergy.

\begin{figure}[t]
    \centering
    \begin{subfigure}[b]{0.49\linewidth}
        \centering
        \includegraphics[width=\linewidth]{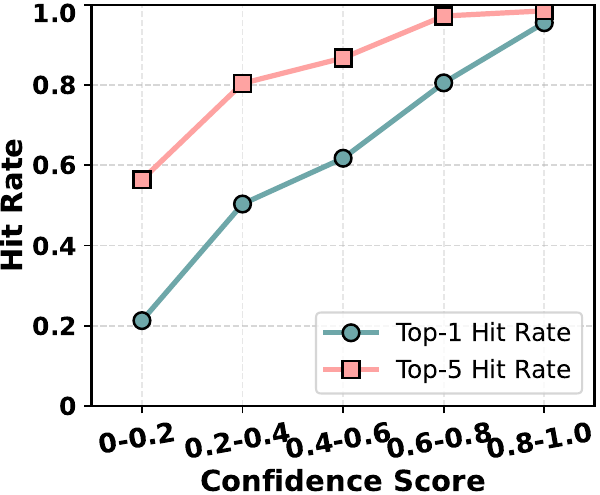}
        \caption{\footnotesize Hit rate vs. confidence score}
    \end{subfigure}
    \hfill
    \begin{subfigure}[b]{0.48\linewidth}
        \centering
        \includegraphics[width=\linewidth]{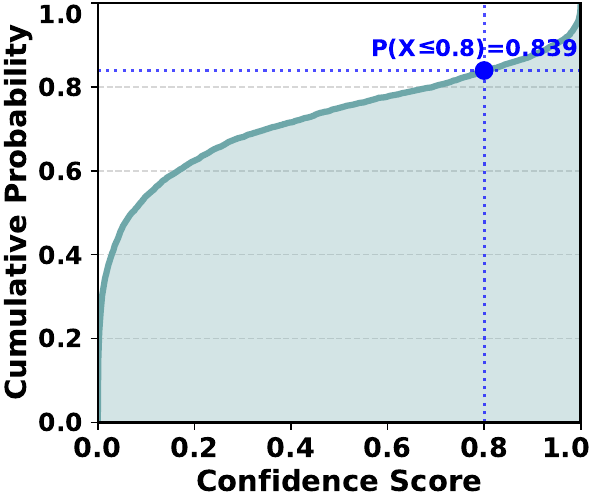}
        \caption{\footnotesize The CDF of confidence score}
    \end{subfigure}
    \caption{
        The accuracy of the SLM’s predictions with respect to the LLM (left). The higher the probability a SLM assigns to a token, the more likely it will match the LLM's prediction. However, the tokens with high probability ($>$0.8) only account for 16.1\% (right).
    }
    \label{fig:confidence}
\end{figure}

\textbf{Opportunity.} LLMs can also indicate its confidence towards prediction via confidence score, but in the token level. As shown in Fig.~\ref{fig:workflow}, the confidence score corresponds to each token's top-1 probability.
Fig. \ref{fig:confidence} (a) demonstrates its effectiveness. The top-1 hit rate measures how often the SLM’s predicted token exactly matches the LLM, while the top-5 hit rate indicates whether the correct token is among the SLM’s five most likely predictions.
When the confidence is high (0.8-1.0), the SLM prediction consistently matches LLM. Notably, the top-5 hit rate remains above 50\% consistently. Therefore, confidence scores are an effective metric to distinguish highly confident tokens to retain locally, and offload low confident tokens to the cloud-based LLM.

\textbf{Limitations of confidence score.} However, SLM are typically ``under‑confident''. In Fig. \ref{fig:confidence} (b), we plot the CDF curve of confidence scores. Only 16.1\% of tokens exceed a confidence score of 0.8, while most fall in the 0–0.4 range.
Furthermore, in a challenging summary task on CNNDM dataset~\cite{cnndm_}, we evaluate the SLM-LLM synergy with a confidence threshold: tokens below this threshold are offloaded to cloud for verification, while those above are accepted by default SLM’s output. As the threshold decreases from 1.0 to 0.8, it does not encounter quality loss. However, decreasing the threshold from 0.8 to 0.7 leads to an unexpected 30\% drop in summary quality. This highlights a limitation of the confidence score: 
confidence works for coarse filtering of highly confident tokens ($\sim 15\%$) to retain locally, but fails to discriminate among the uncertain majority ($\sim 85\%$). Therefore, an additional metric is needed for fine-grained filtering.

\textbf{Importance score.} To address this issue, we introduce the importance score. It is obtained from the attention matrix in the transformer architecture, as depicted in Fig. \ref{fig:workflow}. 
The attention module is defined as:

\begin{equation*}\small
\mathrm{Attention}(Q, K, V) = \mathrm{softmax}\left({QK^{\top}}/{\sqrt{d_k}}\right)V
\end{equation*}

\begin{figure}[t]
    \centering
    \begin{subfigure}[b]{0.49\linewidth}
        \centering
        \includegraphics[width=\linewidth]{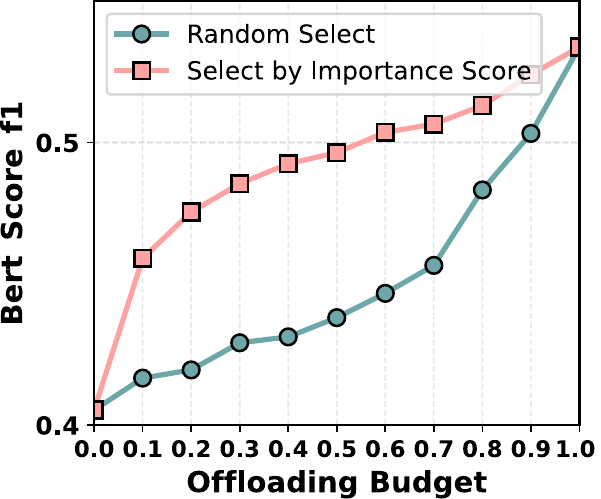}
        \caption{\footnotesize Acc. vs. offloading budget}
    \end{subfigure}
    \hfill
    \begin{subfigure}[b]{0.48\linewidth}
        \centering
        \includegraphics[width=\linewidth]{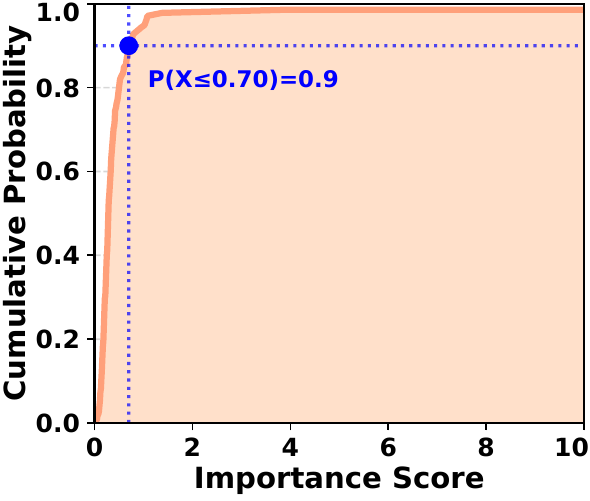}
        \caption{\footnotesize The CDF of importance score}
    \end{subfigure}
    \caption{
        The SLM ranks tokens by importance score and offloads the top n\% to the LLM, achieving sharp quality gains with only a 10–20\% budget (left). The importance scores exhibit a long-tail distribution, where quality-critical tokens are scarce (right).
    }
    \label{fig:importance}
\end{figure}

In particular, we pay attention to the attention matrix $\mathrm{softmax}(QK^{\top}/\sqrt{d_k})$ which captures the token-level ``attention'' paid to one during generation. 
Prior studies on attention sparsity and token pruning~\cite{zhang2023h2o,wang2023tabi} show that column‑wise summation yields an importance score, assigning higher weights to critical tokens. Unlike the confidence score, the importance score enables fine‑grained filtering. Fig.~\ref{fig:importance}(b) further reveals its long‑tail distribution, where high‑importance tokens constitute only a small fraction.

To assess the effectiveness of the importance score, we conduct a measurement study on the CNNDM dataset by offloading the generation of high-importance tokens to the LLM to verify. In Fig.~\ref{fig:importance} (a), a higher BERTScore reflects better generation. Offloading the top 10\% most important tokens yields notable quality gains, and performance matches full LLM generation as the budget increases from 0.3 to 1.0. In contrast, randomly selecting tokens at the same budget (green curve) produces markedly lower quality.

\textbf{Takeaway-1:} 
\textit{Confidence score serves as a coarse-grained filter to locally retain the most confident and certain tokens, while importance score is used for fine-grained selection.}

\subsection{Pipeline Stalls Waste Device Efficiency}\label{motivation: pipeline stalls}

\begin{figure}[t]
    \centering
    \includegraphics[width=0.98\linewidth]{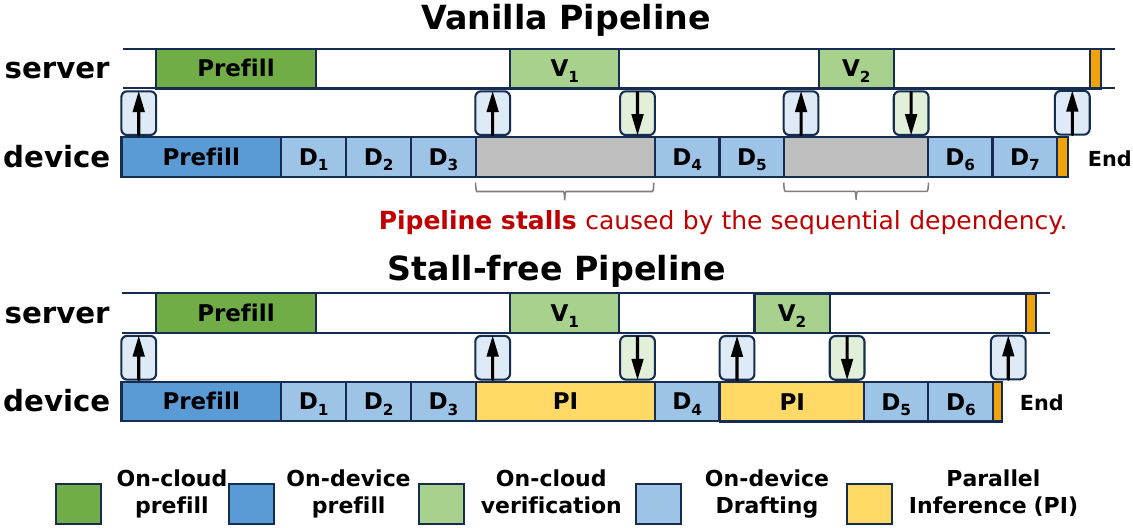}
    \caption{
        The vanilla pipeline (up) shows the on-device pipeline stalls during offloading due to the sequential dependency in auto-regression. Our goal is to mitigate the stalls and unleash the device efficiency.
    }
    \label{fig:pipeline}
\end{figure}

Fig.~\ref{fig:pipeline} illustrates a detailed vanilla pipeline of token-level synergy. Initially, the device performs the prefilling phase, followed by several rounds of generation. When encountering quality-critical tokens as discussed in \S~\ref{motivation: confidence and importance}, the device sends a query to the server for verification. After the uplink communication, cloud-based verification, and downlink communication are completed, the device resumes generation and repeats this process until the end of generation.

\textbf{Challenges.} The vanilla pipeline design suffers from inefficiencies due to pipeline stalls. These stalls, marked in gray in Fig.~\ref{fig:pipeline}, include communication-induced stalls, constrained by network bandwidth, and computation-induced stalls, caused by verification latency. Empirical results reveal that pipeline stalls can constitute a large portion of the total inference time. Under a constrained 50 Kbps bandwidth, communication-induced stalls can account for up to 30\% of the device-side inference time. Meanwhile, when the LLM verification latency reaches 400 ms per verification, computation-induced stalls can consume as much as 50\% of the total inference time. The main cause of stalls is the computation dependency from auto-regression: during draft \& verify in Fig.~\ref{fig:speculative}, the SLM must wait for cloud verification before generating the next token.
Thus, the pipeline stalls significantly harm the pipeline efficiency. 

\textbf{Opportunity.} To address this bottleneck, it is crucial to design a system module that can overlap cloud and device computation, thereby unleashing the parallel potential. The opportunity lies in the output alignment between the SLM and the LLM. As shown in Fig.~\ref{fig:confidence} (a), when the confidence score is high, the SLM often matches the LLM in both top-1 and top-5 predictions. Based on this observation, we can design a module that speculatively predicts both which tokens the LLM may accept and what the likely corrections might be during verification. This enables us to overlap device and cloud pipelines, effectively masking communication and verification delays.

\textbf{Takeaway-2:} 
\textit{There can be a significant pipeline stall resulting from computation dependency, but SLM can overlap stalls and computation, by continuing generation in parallel based on the SLM-LLM output alignment.}

\subsection{Batching Bottleneck in Cloud Serving}\label{motivation:cloud_serving}

The cloud side of our synergistic serving system is responsible for efficiently handling verification requests from multiple devices, with two key goals: (1) \textit{Scalability.} The cloud must support a growing number of requests while keeping latency low. (2) \textit{Compatibility.} The design should minimize gaps from existing cloud-based serving (e.g., vLLM) to leverage established efficiencies and enable seamless deployment together with cloud systems without major changes.

\begin{figure}[t]
    \centering
    \includegraphics[width=0.95\linewidth]{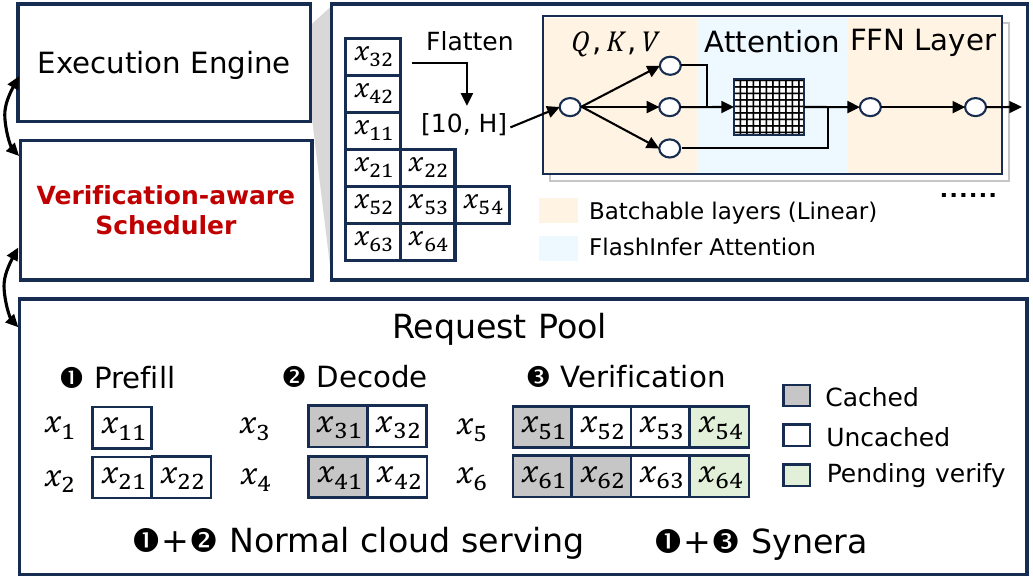}
    \caption{
        The token-level offloading introduces intermittent and non-uniform verification requests (marked with \ding{184}) in the request pool. We aim to handle this heterogeneous request with a verification-aware scheduler to enable scalability and compatibility.
    }
    \label{fig:cloud_serving}
\end{figure}

\textbf{Challenges.} Our proposed token-level offloading introduces verification requests. They are heterogeneous compared to prefill and decode requests, and cannot be processed by normal continuous batching. Fig. \ref{fig:cloud_serving} shows that the verification request comprises cached tokens, uncached tokens, and the pending verify tokens, such as $x_5$ and $x_6$. These requests are characterized by the following properties:
(1) \textit{Intermittency.} During the generation pipeline (Fig. \ref{fig:pipeline}), verification requests may occur at any decoding step whenever quality-critical tokens are encountered.
(2) \textit{Non-uniformity.} Verification requests exhibit significant variability in token length, leading to non-uniform computation. This non-uniformity makes token-level offloading incompatible with existing cloud-based continuous batching schedulers.

\textbf{Opportunity.} Despite the challenges, we identify a key opportunity in reinterpreting verification requests. Specifically, each verification request can be logically decomposed into two components: (1) \textit{The device-accepted token.} This is a sequence of tokens already validated and accepted by the device, but remain uncached by the cloud. It requires to be forwarded to compute KV cache. (2) \textit{The pending verify token}, which also requires a batched forward pass, but followed with a verification as shown in Fig. \ref{fig:speculative}.
Therefore, the two components can be processed similarly to a prefill request, but with an existing KV-cached prefix. We refer to this as \textit{partial prefill}, somehow similar with the chunked prefill discussed in Sarathi-Serve~\cite{agrawal2024taming_sarathiserve}. This insight allows us to preserve the efficiency and throughput benefits of continuous batching, while adapting it to token-level offloading.

\textbf{Takeaway-3:} 
\textit{Token-level offloading creates intermittent and non-uniform requests. They are difficult to handle in normal continuous batching. Treating them as partial prefills can reuse existing batching methods efficiently.}

\section{\name{}}

\subsection{System Overview}
\begin{figure}[t]
    \centering
    \includegraphics[width=0.95\linewidth]{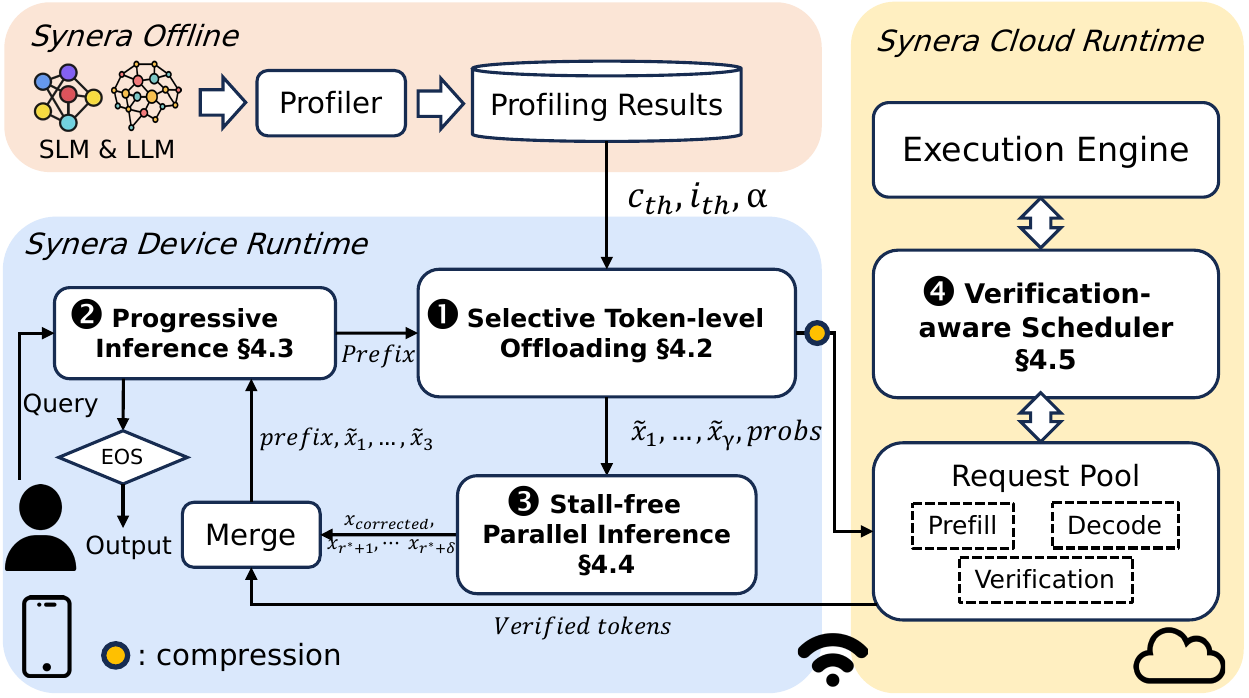}
    \caption{
        The overview of \name.
    }
    \label{fig:system_overview}
\end{figure}

Based on our insights from Section~\ref{motivation}, we design \name{}, an edge-cloud synergistic system with token-level offloading. Fig.~\ref{fig:system_overview} overviews the design. \name~consists of two stages, including the online device and cloud execution runtime, and offline \name-aware profiling. 
Before runtime deployment, \name~conducts offline profiling and prepares profile results (\S~\ref{Implementation}). 
At runtime, the device side conducts model auto-regressive inference with \ding{182} selective token-level offloading (\S~\ref{system design: selective token-level offloading}). \ding{183} Progressive early exit (\S~\ref{system design: early exit}) allows efficient inference with early termination both at the layer and sequence level when confidence and context are sufficient. When encountering quality-critical token chunks, the device runtime asynchronously queries cloud runtime for verification and continues \ding{184} stall-free parallel inference (\S~\ref{system design: stall-free parallel inference}). After receiving the verification request, the \ding{185} verification-aware scheduler (\S~\ref{system design: verification-aware scheduler}) will schedule requests, prefill and verification together, and feed them into the execution engine. The output of verification will return to the device runtime and merge with the output.

\subsection{Selective Token-level Offloading}\label{system design: selective token-level offloading}

Our insights in Section \ref{motivation: confidence and importance} show that confidence scores serve as a coarse-grained filter to retain reliable tokens locally, while importance scores provide a fine-grained, quality-aware selection. We develop a selection method to integrate both, with sampling decisions governed by $P_{imp}$ and $P_{conf}$.

\begin{figure}[t]
    \centering
    \begin{subfigure}[b]{0.48\linewidth}
        \centering
        \includegraphics[width=\linewidth]{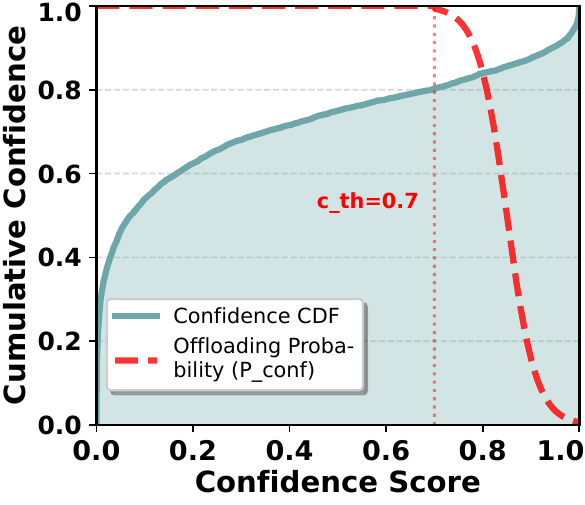}
        \caption{\footnotesize The CDF of confidence score and a visualization of $P_{conf}$.}
    \end{subfigure}
    \hfill
    \begin{subfigure}[b]{0.48\linewidth}
        \centering
        \includegraphics[width=\linewidth]{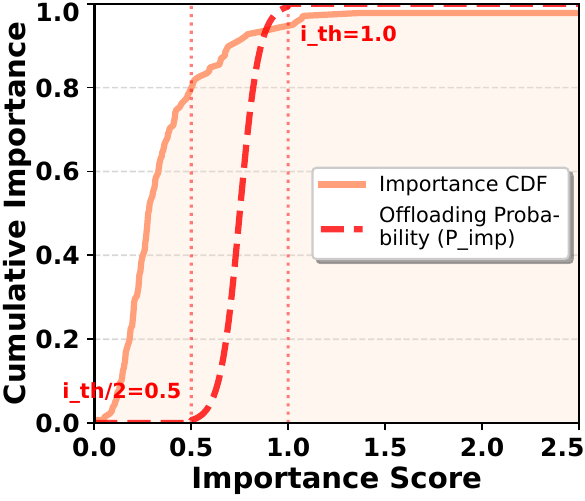}
        \caption{\footnotesize The CDF of importance score and a visualization of $P_{conf}$.}
    \end{subfigure}
    \caption{
        Illustration of confidence and importance dispatching probability, including confidence dispatching probability (left), importance dispatching probability (right), and their corresponding CDF curves.
    }
    \label{fig:p_conf and p_imp}
\end{figure}

\begin{figure}[t]
    \centering
    \includegraphics[width=0.95\linewidth]{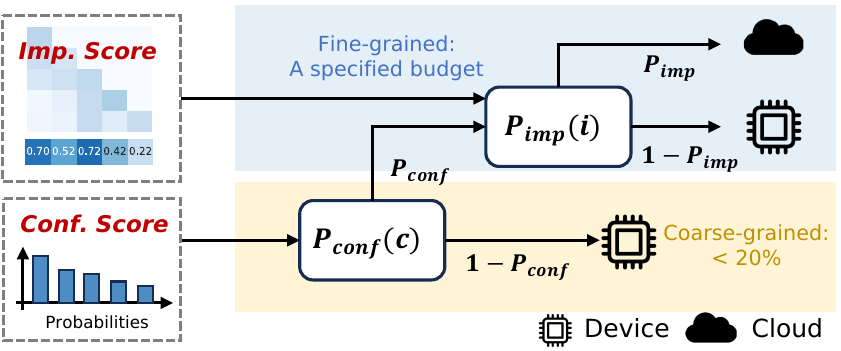}
    \caption{
        An illustration of selective offloading. First, $P_{conf}$ performs a coarse filtering by retaining confident tokens ($<20\%$). Next, $P_{imp}$ makes fine-grained decisions on unconfident tokens ($>80\%$) under a specified budget.
    }\label{fig: put conf. and imp. together}
\end{figure}

\textbf{Confidence score as a coarse-grained filter.} 
First, we obtain the confidence score after forward passes as shown in Fig.~\ref{fig:workflow}. 
We model the dispatch probability for offloading a token chunk to the cloud using a scaled sigmoid function parameterized by a threshold $c_{th}$:

\begin{equation*}\label{equation: probablity}\small
P_{conf}(c) = \begin{cases} 
\frac{1}{1 + \exp(k \cdot norm(c))} & c > c_{th} \\
1 & c \leq c_{th}
\end{cases}
\end{equation*}

where $c$ denotes the average confidence score over draft tokens, and it is normalized by $norm(c) = \frac{c - c_{th}}{1 - c_{th}} - \frac{1}{2}$. $c_{th}$ represents the threshold (0.7-1.0 typically), and $k$ controls the steepness of the sigmoid curve. The hyper-parameters are $c_{th}$ and $k$. We set $k=10$ for a moderate slope, avoiding the instability caused by a high $k$ and the lax filtering resulting from a low $k$ in confidence thresholding. During the offline profiling (\S~\ref{Implementation}), $c_{th}$ is profiled to achieve the best coarse-grained filtering. Fig.~\ref{fig:p_conf and p_imp} (a) visualizes the $P_{conf}$ curve, and Fig.~\ref{fig: put conf. and imp. together} shows the offloading decision process, where the confidence score enables early filtering of highly confident tokens. This approach effectively eliminates the most unnecessary tokens from offloading. At runtime, $c_{th}$ is fixed. 

\textbf{Importance score as a fine-grained filter.}
The importance score enables finer, quality-aware filtering. We also model the probability of offloading by importance using a scaled sigmoid function, which is given by:
\begin{equation*}\label{equation: importance}\small
P_{imp}(i) = \begin{cases} 
0 & i \leq \frac{i_{th}}{2} \\
1 & i > i_{th} \\
\frac{1}{1 + \exp(\theta \cdot norm(i))} & \text{otherwise}
\end{cases}
\end{equation*}
where $i$ denotes the average importance score across draft tokens during generation, and it is normalized by $norm(i) = \frac{i - i_{th}/2}{i_{th}/2} - \frac{1}{2}$. Different with confidence score, thresholds $\tfrac{i_{th}}{2}$ and $i_{th}$ mark lower and upper bounds. This equation yields three tiers: 
tokens with importance scores $\leq \tfrac{i_{th}}{2}$ stay local, $> i_{th}$ are offloaded, and intermediate scores follow a sigmoid distribution with slope $\theta$ ($\theta < 0$). We use $\theta=-10$ for a moderate trade-off between steep and flat curves. We visualize $P_{imp}(i)$ in Fig.~\ref{fig:p_conf and p_imp} (b) and the positioning of importance score is in Fig~\ref{fig: put conf. and imp. together}. The importance score performs fine-grained selection after the confidence score. At runtime, $i_{th}$ is tunable as the offloading budget with quality–latencytrade-offs.

\textbf{Compression before transmission.} To conduct verification, the cloud-based LLM needs both the draft tokens and corresponding probability distribution from the on-device SLM, i.e., $p(x | prefix), \ldots, p(x|prefix, \tilde{x}_1, \ldots, \tilde{x}_{\gamma})$ for verification, according to the ``draft \& verify'' process in Fig.~\ref{fig:speculative}. Each probability distribution contains tens of thousands of float values (e.g., 32,000 for Llama-2 series). This represents a large data volume, and takes over 50~ms for transmission under a typical 10~Mbps bandwidth. We compress the distribution by transmitting the most probable probabilities according to the actual sampling method supposed to use, i.e., top-1 (greedy), top-k, or top-p sampling. 
This approach is based on the already-intended sampling strategy, but can reduce the data size by over 99.5\%, while maintaining lossless impact on verification.

\subsection{Progressive Early Exit Inference}\label{system design: early exit}

\name{} adopts a progressive inference strategy that is to reduce the redundancy during offloading decision, and it operates at two different granularities:  
(1) \textit{Layer-wise early exit}, whose primary goal is to produce the earliest possible confidence/importance signals for selective offloading; and  
(2) \textit{Sequence-wise early exit}, which disables unnecessary cloud verification near the tail of generation.

\textbf{Layer-wise early exit.}  
Layer-wise early exit accelerates the computation of the offloading decision metrics $P_{conf}$ and $P_{imp}$.
Selective offloading decides when to query the cloud based on confidence and importance score. Our observation shows that the two values, especially the confidence score, can stabilize well before the final transformer layer. Therefore, running all layers just to obtain virtually identical $(P_{conf},P_{imp})$ is wasteful in both latency and energy, and delays the moment we can launch an offload request. Therefore, we adopt layer-wise early exit. At each transformer layer $l$, we compute a margin score~\cite{edgefm} from the hidden state $h_l$ using the difference between the top-1 and top-2 probabilities. When this score exceeds a threshold, the model stops further computation and outputs the current confidence score, importance score, and token. We conservatively allow exit only in the last 25\% of layers, and adopt a threshold of 0.7. The impact of the threshold will be analyzed in Section \ref{section: Sensitivity Analysis}.

\textbf{Sequence-wise early exit}. 
This module targets redundancy in token-level offloading, which primarily arises during long-text generation. 
It is motivated by the observation that in the later stages of sequence generation (e.g., after $0.8 \times \text{max\_len}$), the overall SLM token trajectory is informative enough and well-established. At this phase, the SLM can reliably continue generation without further offloading to the LLM for verification. Concretely, we disable token-level offloading when the current generation step $t$ satisfies $t > \gamma \cdot \text{max\_len}$ (with $\gamma$ empirically set to 0.8). We observe that this module can reduce unnecessary communication, without degrading generation quality. 

\subsection{Stall-free Parallel Inference}\label{system design: stall-free parallel inference}

Given the insight from Section~\ref{motivation: pipeline stalls}, this approach consists of two key steps: (1) \textit{rejection position prediction}, which identifies where the local and cloud model outputs are likely to diverge, and (2) \textit{parallel inference}, where the device generates tokens in parallel with cloud computation to maximize efficiency.

\textbf{Rejection position prediction}. 
Our insight is that the rejection position is governed by both the expected number of accepted tokens, modeled as a capped geometric variable~\cite{leviathan2023fast_google_speculative}, and each token’s confidence level, which reflects its acceptance likelihood. Therefore, we leverage a confidence-adjusted geometric distribution to sample the probable rejection position. 
Given a draft token chunk of $\gamma$ tokens ${x_1, x_2, \ldots, x_\gamma}$ with corresponding confidence scores ${c_1, c_2, \ldots, c_\gamma}$, we first establish a base rejection probability distribution following the capped geometric series 
$P_{\text{base}}(r = t) = (1-\alpha)\alpha^t\ \text{if}\ t < \gamma-1,\ \text{else}\ \alpha^\gamma$.
Here, $\alpha$ denotes the expected per-token acceptance probability, profiled offline and discussed in Section \ref{Implementation}, and $r$ is the rejection position. We refine and adjust this distribution by modulating it with each token’s confidence score: $P_{\text{adj}}(r = t) = P_{\text{base}}(r = t) \cdot (1 - c_t)$.
where higher confidence scores $c_t$ reduce the rejection probability at position $t$, because a higher confidence score by SLM demonstrates a lower rejection rate by LLM, as shown in Fig. \ref{fig:confidence} (a). The final prediction distribution is obtained by normalizing: $P(r = t) = \frac{P_{\text{adj}}(r = t)}{\sum_{i=0}^{\gamma-1} P_{\text{adj}}(r = i)}$. We sample a single rejection position $r^*$ from this distribution. As the core of parallel inference, we particularly discuss the hit rate of rejection position prediction in Section~\ref{evaluation: ablation study}.

\textbf{Parallel Inference (PI)}. Upon identifying the predicted rejection position $r^*$, the device launches targeted parallel inference while the cloud is verifying the draft. PI constructs a corrected prefix by retaining tokens up to position $r^*-1$ and replacing the rejected token at position $r^*$ with an alternative sampled from the top-3 candidates of the local model's probability distribution. The top-3 candidates balance diversity and efficiency, as most mispredicted tokens already lie within this range. PI then continues generation from this corrected prefix for up to $\delta$ additional tokens, i.e., $x_{\text{corrected}} \rightarrow [x_{\text{corrected}}, x_{r^*+1}, x_{r^*+2}, \ldots, x_{r^*+\delta}]$. To \textit{merge} the local and cloud results, \name{} compares the actual rejection position $r_{\text{cloud}}$ with the predicted $r^*$. If they match, the PI-generated tokens are adopted, minimizing pipeline stalls. Otherwise, generation resumes from the cloud’s corrected prefix. This approach speculatively reduces pipeline stalls.

\begin{algorithm}[t]
\caption{Verification-aware Scheduler}\label{algorithm:scheduler}
Initialize current active batch of requests $\text{B} \gets \emptyset$\;

\While{True}{
    $\text{B\_prefill} \gets \emptyset$, $\text{B\_verify} \gets \emptyset$\;
    $\text{R\_new} \gets \text{get\_next\_prefill\_request()}$\;
    \While{$\text{R\_new} \neq \emptyset$}{\label{line:prefill-start}
        $\text{B\_prefill} \gets \text{B\_prefill} \cup \text{R\_new}$\;
        $\text{R\_new} \gets \text{get\_next\_prefill\_request()}$\;
    }
    
    \If{$\text{B\_prefill} \neq \emptyset$}{
        $\text{execute\_prefill}(\text{B\_prefill})$\;
        $\text{B} \gets \text{B\_prefill} \cup \text{B}$\;
        \textbf{break}\;
    }\label{line:prefill-end}
    
    $\text{R\_verify} \gets \text{get\_next\_verification\_request()}$\;\label{line:verify-start}
    $\text{L\_uncached} \gets \emptyset$\;
    
    \While{$\text{R\_verify} \neq \emptyset$}{
        $\text{B\_verify} \gets \text{B\_verify} \cup \text{R\_verify}$\;
        $\text{R\_verify} \gets \text{get\_next\_verification\_request()}$\;
        $\text{u} \gets \text{get\_num\_uncached(R\_verify)}$\;
        $\text{L\_uncached} \gets \text{L\_uncached} \cup \text{u}$\;
    }
    
    \If{$\text{B\_verify} \neq \emptyset$}{
        $\text{execute\_partialprefill}(\text{B\_verify}, \text{L\_uncached})$\; 
        $\text{B} \gets \text{B} \cup \text{B\_verify}$\;
    }\label{line:verify-end}
    
    $\text{B} \gets \text{remove\_completed}(\text{B})$\;
}
\end{algorithm}

\subsection{Verification-aware Scheduler}\label{system design: verification-aware scheduler}

Our insights in Section~\ref{motivation:cloud_serving} show that verification requests contain the uncached tokens and pending-verify tokens. They resemble prefill requests but have cached prefixes. To batch verification requests, it is impractical to simply batch these \textit{partial prefill} with normal prefills together. However, prefills usually contain thousands of tokens, while verification requests involve only tens of tokens; mixing them causes GPU pipeline bubbles and high TBT latency~\cite{agrawal2024taming_sarathiserve}.

To address this, we isolate prefill and verification requests into separate iterations. Building on vLLM’s continuous batching, we design a verification-aware scheduler, see Algorithm~\ref{algorithm:scheduler}. In each scheduling iteration, prefill requests are prioritized (lines~\ref{line:prefill-start}–\ref{line:prefill-end}); if none exist, the scheduler processes verification requests (lines~\ref{line:verify-start}–\ref{line:verify-end}). Following Sarathi-Serve~\cite{agrawal2024taming_sarathiserve}, we further apply chunked partial prefill, segmenting them into fixed-length chunks of size 32 to maximize hardware efficiency and kernel compatibility.

Pending verification requests and their uncached tokens are then grouped into a batch $\text{B\_verify}$, which is executed using $\text{execute\_partial\_prefill()}$. Then, as we show in Fig.~\ref{fig:cloud_serving}, the engine flattens tensors across requests before feeding them into the transformer. After the execution engine has finished the forward pass, we conduct verification using the engine's output $q(x | prefix), \ldots, q(x|prefix, \tilde{x}_1, \ldots, \tilde{x}_{\gamma})$. Treating verification requests as partial prefill keeps the scheduler fully compatible with continuous batching frameworks, ensuring seamless integration and efficient handling of intermittent, non-uniform workloads.

\begin{table}[t]
\centering
\resizebox{0.96\linewidth}{!}{
\begin{tabular}{llccc}
\toprule
\makecell{\shortstack[l]{Dataset}} & \makecell{\shortstack[l]{Task}} & \makecell{\shortstack[l]{Metric}} & \makecell{\shortstack[l]{Shot}} \\
\midrule
CSQA~\cite{talmor2018commonsenseqa} & General knowledge QA & Accuracy & 5-shot \\
SST2~\cite{socher2013recursive_sst2} & Sentiment analysis & Accuracy & 5-shot \\
CNN/DM~\cite{cnndm_} & Articles summary & Rouge-1 & / \\
XSum~\cite{Narayan2018DontGM_xsum} & Articles summary & Rouge-1 & / \\
\midrule
LLQA~\cite{llqa} & Mobile daily logger QA & Accuracy & / \\
HeySQuAD~\cite{wu2023heysquad} & Mobile speech answering & Rouge-1 & 5-shot \\
SensorQA~\cite{reichman2025sensorqa} & Mobile sensor QA & Rouge-1 & 5-shot \\
\bottomrule
\end{tabular}
}
\caption{Overview of the evaluation datasets in our experiments. LLQA, HeySQuAD and SensorQA are specially selected for mobile-centric tasks.}\label{Table: datasets}

\end{table}

\section{Implementation}\label{Implementation}

\textbf{Runtime implementation.} At runtime, \name{} have two components: 
(1) \textit{Device Runtime:} 
We implement \name{} device runtime prototypes on two mobile and edge devices: a Jetson AGX Orin (32 GB RAM) running a $\sim$3k LoC atop Transformer~\cite{transformers_lib}, and a Google Pixel 7 smartphone using mllm ~\cite{yi2023mllm} as the backend. Jetson power modes are tuned to match SLM and platform characteristics. The two devices span diverse scenarios and reflect the practical use of LLMs on mobile and edge devices. 
(2) \textit{Cloud Runtime:} 
\name~cloud runtime is on a cloud server with 8 A6000 GPUs (48GB memory) running Ubuntu 20.04. It is built with $\sim$2k LOC using python atop Sarathi-Serve~\cite{agrawal2024taming_sarathiserve}. Sarathi-Serve, forked from vLLM, is a high throughput and low-latency cloud-based LLM serving framework via continuous batching.
While \name{} uses Sarathi-Serve, it can extend to any framework with continuous batching, e.g., SGLang~\cite{sglang}.

\textbf{\name~offline profiling.} 
For each SLM-LLM model pair, we need to profile parameters of each \name{} modules, i.e., cut-off confidence score $c_{th}$, cut-off importance score $i_{th}$ and probability $\alpha$ that a drafted token is accepted by the LLM, to identify the optimal parameter settings. 
To profile these parameters, we perform inferences with all tokens offloaded from SLM to LLM. We obtain the cut-off $c_{th}$ by collecting the fully accepted token chunks and recording their average confidence score during each offloading. Given that the budget is tunable, we pre-collect the distribution of importance scores for all token chunks. For a specific budget, we set the cut-off $i_{th}$ at the corresponding percentile of this distribution. We calibrate the acceptance probability $\alpha$ using the expectation of the capped geometric distribution $E(\# \text{generated tokens}) = \frac{1-\alpha^{\gamma+1}}{1-\alpha}$~\cite{leviathan2023fast_google_speculative},
where $\gamma$ denotes the draft length. We set $\gamma = 4$ by default, since it is a moderate offloading token length granularity.

\section{Evaluation}

\subsection{Experimental Setup}\label{evaluation: exp setup}

\textbf{Datasets and models.}
Table~\ref{Table: datasets} lists the datasets from real-world scenarios, covering key LLM evaluation tasks such as question answering and summarization.
We further include mobile-centric datasets (i.e., LLQA, HeySQuAD, SensorQA) to emphasize evaluation in interactive mobile contexts such as daily logger QA and sensor understanding.
Table~\ref{Table: models} presents 9 representative SLMs and LLMs. We report their accuracy on CommonsenseQA using HuggingFace implementations.

\begin{table}[t]
\centering
\resizebox{0.95\linewidth}{!}{
\begin{tabular}{lcccc}
\toprule
\makecell{\shortstack[l]{Model (Acronym)}}  & \makecell{\shortstack[l]{Parameters (B)}} & \makecell{\shortstack[l]{Accuracy (\%)}} & 
\makecell{\shortstack[l]{\textit{$Pf$}}} & Deployment\\
\midrule
Llama-70B~\cite{touvron2023llama}         & 70    & 76.82      &  1 & Cloud\\
Llama-13B~\cite{touvron2023llama}         & 13    & 66.99      &  6 & Cloud\\
Llama-7B~\cite{touvron2023llama}          & 7     & 62.49   &  13 & Device\\
Llama-1.1B~\cite{zhang2024tinyllama}        & 1.1   & 19.66   &  86 & Device\\
Llama-160M~\cite{miao2024specinfer}         & 0.16  & 18.85   &  558 & Device\\
\midrule
Yi-34B~\cite{ai2024yi}              & 34    &  84.28     &  2 & Cloud\\
Yi-6B~\cite{ai2024yi}               & 6     &  75.27    &  15 & Device\\
\midrule
Falcon-40B~\cite{almazrouei2023falcon}          & 40    &   70.19    &  2 & Cloud\\
Falcon-7B~\cite{almazrouei2023falcon}           & 7     &   21.46    &  13 & Device\\
\bottomrule
\end{tabular}
}
\caption{
The LLMs and SLMs used in the evaluation.
}
\label{Table: models}

\end{table}

\begin{table*}[!htbp]
\setlength{\abovecaptionskip}{2pt}
\setlength{\belowcaptionskip}{2pt}
\centering
\resizebox{0.9\textwidth}{!}{
\begin{tabular}{l|ccccccccc}
\toprule
\multirow{2}{*}{\shortstack[l]{Model Pair}}  & \textbf{Dataset} & \textbf{CNNDM} & \textbf{XSum} & \textbf{SensorQA} & \textbf{HeySQuAD} & \textbf{CSQA} & \textbf{SST2} &  \textbf{LLQA}    \\
& Metrics & Rouge-1$\uparrow$ & Rouge-1$\uparrow$ & Rouge-1$\uparrow$ & Rouge-1$\uparrow$ & Acc$\uparrow$ & Acc$\uparrow$ & Acc$\uparrow$  \\

\midrule

\multirow{4}{*}{\shortstack[l]{Llama-160M \\\& Llama-13B}}
& Edge-centric  & 3.23 & 4.78 & 3.98 & 3.24 & 18.85 & 53.78 & 23.80 \\
& EdgeFM-LLM    & 8.58 & 5.33 & 12.35 & 10.59 & 21.08 & 54.26 & 37.84 \\
& Hybrid        & 14.61 & 7.32 & 11.70 & 12.37 & 22.43 & 58.73 & 45.32 \\
& \textbf{\name}& \textbf{17.67} & \textbf{8.91} & \textbf{13.77} & \textbf{17.74} & \textbf{31.22}  & \textbf{69.20} & \textbf{55.03} \\

\midrule

\multirow{4}{*}{\shortstack[l]{Llama-1.1B \\\& Llama-13B}}
& Edge-centric   & 18.29 & 10.95 & 18.84 & 19.11 & 19.66 & 80.16 & 26.23 \\
& EdgeFM-LLM   & 20.63 & 13.56 & 27.67 & 30.89 & 30.26 & 82.13 & 39.98  \\
& Hybrid   & 20.85 & 12.44 & 26.36 & 31.44 & 30.55 & 81.41 & 39.92 \\
& \textbf{\name} & \textbf{21.87} & \textbf{16.04} & \textbf{32.70} & \textbf{36.35} &  \textbf{30.94} & \textbf{84.76} & \textbf{42.97} \\

\midrule

\multirow{4}{*}{\shortstack[l]{Llama-7B \\\& Llama-70B}} 
& Edge-centric   & 23.51 & 21.24 & 39.63 & 66.55 & 62.57 & 87.16 & 82.02 \\
& EdgeFM-LLM     & 23.66 & 22.99 & 42.33 & 69.15 & 64.60 & 89.26 & 85.70 \\
& Hybrid         & 25.23 & 25.77 & 41.01 & 71.98 & 70.90 & 91.23 & 92.62 \\
& \textbf{\name} & \textbf{27.11} & \textbf{30.07} & \textbf{47.88} & \textbf{72.35} & \textbf{72.08} & \textbf{91.98} & \textbf{95.98} \\
\bottomrule
\end{tabular}
}
\caption{Comparison of end-to-end generation quality on various datasets. We take three model pairs and report the corresponding quality metrics on selected baselines.}\label{table: end to end quality}
\vspace{-15pt}
\end{table*}

\textbf{Baselines.}
We implement the following 4 baselines:
(1)
\textit{Edge-centric}. This baseline performs all LLM inference locally on devices, without reliance on cloud resources.
(2)
\textit{Cloud-centric}. This method offloads all user queries to the cloud. We deploy larger LLMs as listed in Table \ref{Table: models} using the Sarathi-Serve~\cite{agrawal2024taming_sarathiserve} inference engine.
(3)
\textit{Hybrid}~\cite{native_hybrid_renju_zixuhao}, which is an SLM and LLM hybrid inference framework by offloading uncertain tokens to the cloud-based LLM. We implement this hybrid approach to compare within the SLM-LLM synergistic inference paradigm. 
(4)
\textit{EdgeFM-LLM}. EdgeFM~\cite{edgefm} is a device-cloud framework that identifies hard samples for cloud offloading. Since the original method is designed for discriminative models, we adapt EdgeFM to generative LLM by offloading high-perplexity (PPL) samples. We refer to this adapted variant as \textit{EdgeFM-LLM}. This enables a comparison between input-level and our token-level offloading in the context of LLM serving.

\textbf{Metrics}. Our evaluation employs different performance metrics depending on the dataset, as detailed in Table \ref{Table: datasets}. Rouge-1 scores are reported on a percentage scale (0–100\%) for clarity. For latency, we use the average time between tokens (TBT), which is crucial for user experience; we do not consider time to first token (TTFT) since \name{} does not target its optimization. To evaluate cloud serving costs, we use the packing factor ($Pf$), following previous work~\cite{wang2023tabi,gunasekaran_cocktail_nodate}. The packing factor represents the number of concurrent models that can run on a GPU cluster without causing more than 10\% latency degradation, serving as a proxy for unit cost. Table \ref{Table: models} reports packing factors for the LLMs evaluated. The $Pf$ is normalized by Llama-2-70B. The estimated cloud serving cost $c$ is computed as $c = \frac{1}{Pf} \times T \times W$, where $\frac{1}{Pf}$ is the model cost, $T$ is the average TBT, and $W$ is the average percentage of tokens generated for each dataset.

\subsection{Overall Performance}\label{evaluation: end_to_end_performance}

\textbf{Generation quality.} Table \ref{table: end to end quality} presents the generation quality of various SLM-LLM model pairs across multiple datasets. \name{} achieves substantial quality improvements, ranging from 1.2$\times$ to 5.47$\times$, over edge-centric baselines. Furthermore, \name{} consistently outperforms synergistic serving approaches (EdgeFM-LLM and Hybrid), which utilize sample-level and token-level offloading policies. Notably, the benefits of \name{} are most pronounced when there is a substantial capability gap between the SLM and LLM, such as between Llama-160M and Llama-13B.  
Notably, \name{} enables Llama-160M to reach generation quality comparable to a Llama-1.1B baseline, highlighting its effectiveness in boosting SLMs via LLM verification and unleashing the usability of tiny SLM.

 \begin{figure*}[!htbp] 
    \centering 
     \includegraphics[width=0.95\linewidth]{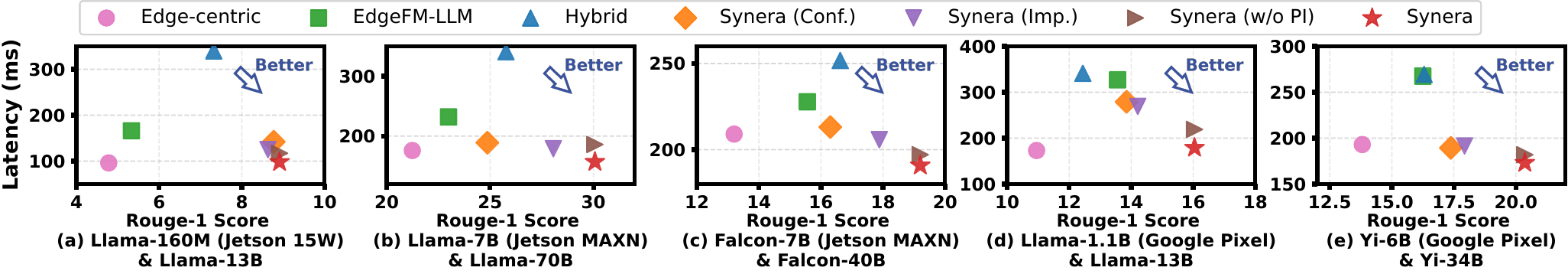}
     \vspace{-8pt}
    \caption{ 
         End-to-end Latency and Generation quality results on XSum dataset. 
    }
    \vspace{-8pt}
    \label{fig:end-to-end-latency} 
\end{figure*} 

\begin{figure*}[!htbp] 
    \centering 
    \includegraphics[width=0.94\linewidth]{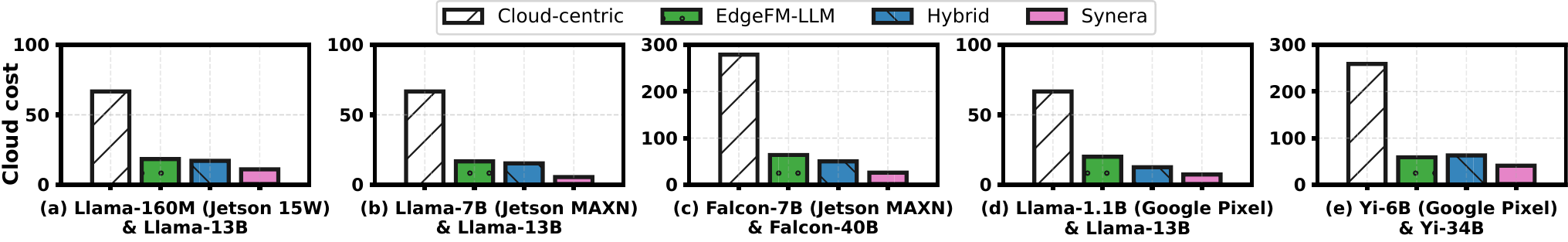} 
     \vspace{-8pt}
    \caption{ 
         End-to-end estimated cloud serving cost on XSum dataset. 
    }
    \label{fig:end-to-end-serving-cost} 
    \vspace{-10pt}
\end{figure*} 

\begin{figure}[t] 
    \centering 
    \includegraphics[width=0.94\linewidth]{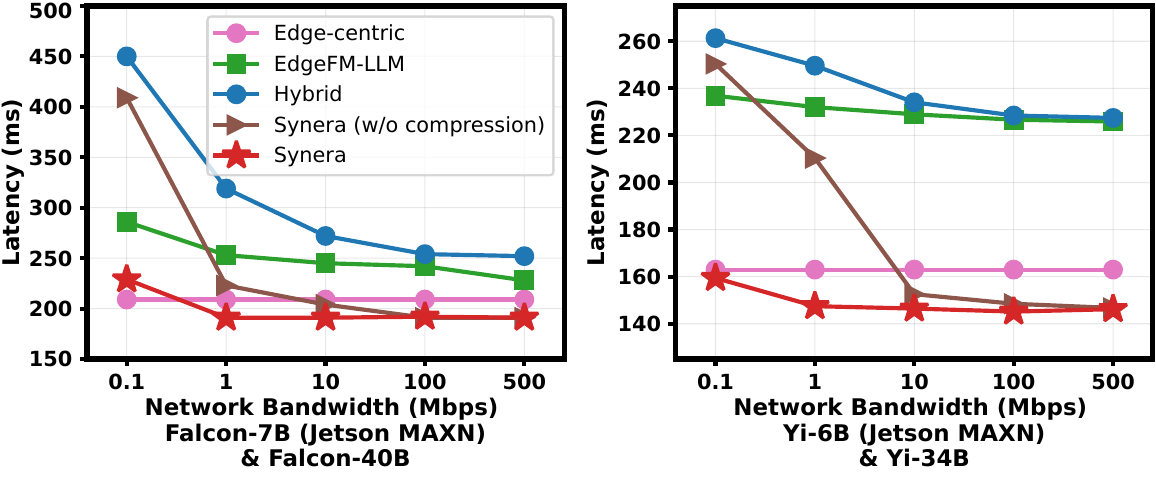 }
    \vspace{-10pt}
    \caption{
        The impact of bandwidth on \name{}. \name{} shows high resilience to various network conditions.  
    }
    
    \label{fig:impact of bandwidth}
\end{figure}

\textbf{Latency.} Figure \ref{fig:end-to-end-latency} shows the end-to-end latency results across five configurations, each comprising an SLM, a device (with energy mode for Jetson Orin), and a cloud-based LLM. \name{} significantly improves generation quality without introducing additional latency compared to edge-centric. Furthermore, \name{} reduces average latency by 32–41\% compared to EdgeFM-LLM, and by 35–71\% compared to Hybrid. While Hybrid also adopts token-level synergy, \name{} achieves superior performance through an optimized offloading policy and efficient pipeline execution. 

\textbf{Estimated cloud serving cost.}
Fig.~\ref{fig:end-to-end-serving-cost} shows cloud serving costs under five deployment configurations. \name{} significantly cuts cloud serving cost by 88\% versus the cloud‑centric approach by offloading only a small fraction of token chunks rather than full inference. Beyond this, it further outperforms all baseline methods, reducing cost by 52\% over EdgeFM‑LLM and 45\% over Hybrid. These results highlight that \name{} enables highly cost-efficient deployment by minimizing reliance on expensive cloud resources.
 
\textbf{Impact of bandwidth.} 
Fig.~\ref{fig:impact of bandwidth} shows latency under varying network bandwidths. \name{} consistently outperforms baselines and sustains low latency even under severe network constraints (0.1~Mbps)
This bandwidth-resilience is attributed to two key factors: (1) token-level synergy, which reduces the amount of data that needs to be offloaded, and (2) optimized data transmission through compression, which effectively mitigates the impact of large data volume.

\subsection{Trade-offs when Budget Changes} \label{evaluation:trade-off}

\begin{figure}[t] 
    \centering 
    \includegraphics[width=0.97\linewidth]{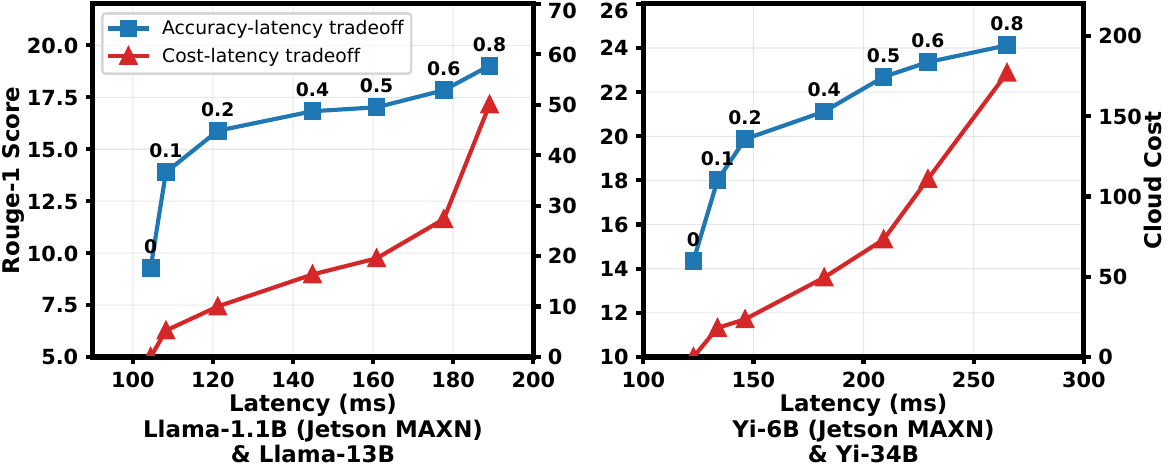 }
    \vspace{-10pt}
    \caption{ 
        The trade-offs between generation quality, cloud cost and latency. The marked numbers represent the offloading budget.
    }
    \label{fig:trade-off} 
\end{figure}

\textbf{Accuracy–Latency Trade-off.} \name{} demonstrates a clear trade-off between generation quality and latency. As depicted in Fig.~\ref{fig:trade-off}, an offloading budget of approximately 0.2 achieves a favorable balance, simultaneously ensuring low latency, competitive accuracy, and limited reliance on cloud resources. Increasing the offloading budget from 0 to 0.8 consistently improves generation quality. This is because a larger offloading budget allows the on-device SLM to leverage more frequent verification from the cloud-based LLM. 

\textbf{Accuracy-Cost Trade-off.} \name{} also highlights a trade-off between generation accuracy and cloud serving cost. Within the initial range (budget 0 to 0.2), generation quality increases steeply while the additional cloud cost remains negligible. This demonstrates that substantial accuracy gains can be attained with minimal expenditure. Nonetheless, this parameter is not fixed; it can be flexibly tuned according to application-specific requirements and resource constraints.

\subsection{Scalability}\label{evaluation: scalability}

\begin{figure}[t]
    \centering 
    \includegraphics[width=0.7\linewidth]{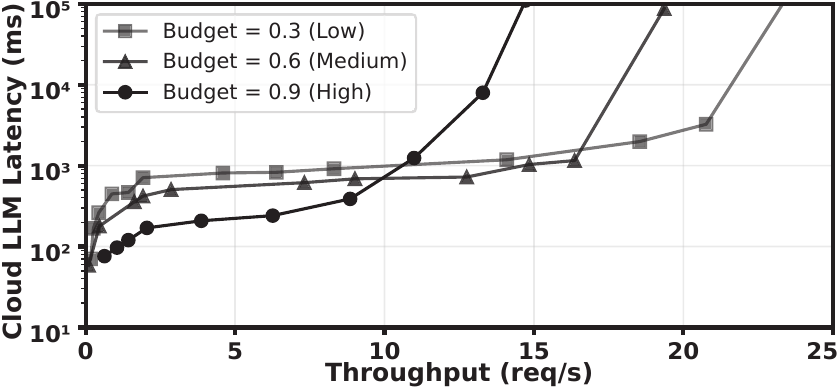}
    \vspace{-8pt}
    \caption{ 
        Latency and throughput with three offloading budgets. Low budget indicates higher scalability. 
    } 
    \label{fig:Scalability} 
\end{figure}

We evaluate the scalability of the \name{} cloud runtime, focusing on the verification-aware scheduler as request load grows. In Fig.~\ref{fig:Scalability}, we compare three offloading budgets: low (0.3), medium (0.6), and high (0.9). Higher budgets issue more frequent device queries to the cloud LLM. The y-axis shows verification latency per request.
Results indicate that latency remains stable while throughput is below thresholds of about 10, 15, and 20 reqs/s for budgets of 0.3, 0.6, and 0.9, respectively. Beyond these points, latency rises sharply.
Lower budgets are more resilient under high throughput, since they issue fewer verification queries. In practice, the system operates in the low-budget setting. Overall, the \name{} runtime scales efficiently with growing request numbers, while the scheduler enables resilience under different request load. 

\subsection{Ablation Study}\label{evaluation: ablation study}

\textbf{Comparison: $P_{conf}$ or $P_{imp}$}. In Fig. \ref{fig:end-to-end-latency}, \name{} (Conf.) and \name{} (Imp.) represent variants of \name{} with only the $P_{conf}$ or $P_{imp}$, instead of using both as we show in Fig. \ref{fig: put conf. and imp. together}. The results in Fig. \ref{fig:end-to-end-latency} demonstrates that the dual-metric offloading strategy in \name{} significantly reduces latency and further enhances quality compared to single-metric variants.

\textbf{Comparison: w/ and w/o stall-free parallel inference.}
\name{} (w/o PI) denotes the variant without parallel inference. As shown in Fig. \ref{fig:end-to-end-latency}, incorporating parallel inference reduces latency compared to the vanilla pipeline. This improvement stems from the rejection position prediction module, which achieves a substantial hit rate in our measurements (38\% for Llama-160M\&13B and 31\% for Llama-1.1B\&13B).

\textbf{Comparison: w/ and w/o compression.}
In Fig. \ref{fig:impact of bandwidth}, \name{} (w/o compression) removes the compression before transmission, as discussed in Section \ref{system design: selective token-level offloading}. The results in low bandwidth, i.e., 0.1 and 1 Mbps, demonstrate the necessity of compression in taming the network latency.

\subsection{Sensitivity Analysis}\label{section: Sensitivity Analysis}
In Fig. \ref{fig:early-exit}, we vary \name{}’s layer-wise early exit threshold from 0.0 (almost certain early exit) to 1.0 (almost no early exit). Threshold set to 0.6 or 0.8 is optimal, as the accuracy (measured by Rouge-1 score) remains nearly unchanged compared to a threshold of 1.0, while latency is reduced by 20\%. This trend is consistent across different model pairs in our experiments, highlighting the effectiveness.

\subsection{System Overhead}\label{evaluation: overhead}

Table~\ref{table: overheads_device} reports two metrics: scheduling latency per token (wall-clock time for $P_{conf}$ and $P_{imp}$) and energy per token (from tegrastats on Jetson Orin). Compared to the edge-centric baseline, \name{} adds negligible runtime overhead: latency is always $<0.5$~ms per token, and energy differs by only -0.03~J, showing no significant extra overhead.

\begin{figure}[!t]
\centering
\begin{minipage}[b]{0.50\linewidth}
    \centering
    \includegraphics[width=\linewidth]{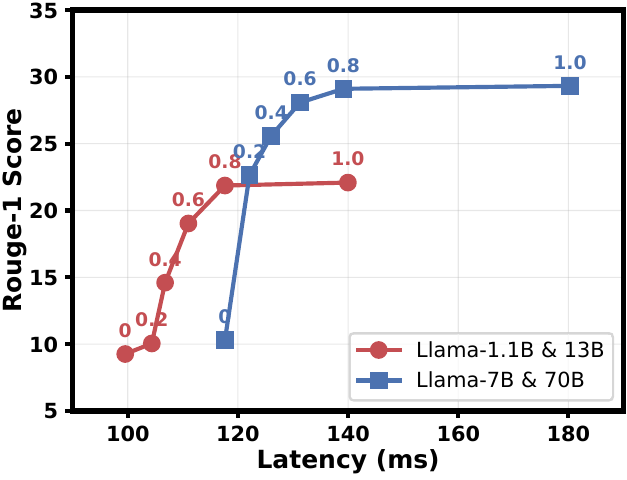}
    \vspace{-20pt}
    \caption{\footnotesize Ablation study of early exit on CNNDM. Marked numbers indicate exit thresholds.}
    \label{fig:early-exit} 
\end{minipage}
\hfill
\begin{minipage}[b]{0.48\linewidth}
    \centering
    \includegraphics[width=\linewidth]{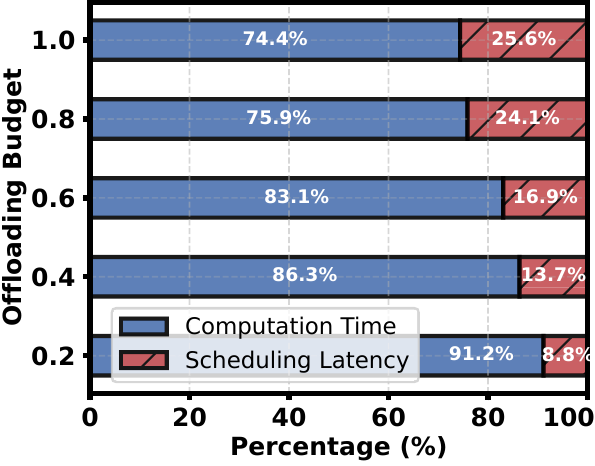}
    \vspace{-20pt}
    \caption{\footnotesize \name{} cloud runtime latency overheads across different offloading budgets.}
    \label{fig:cloud runtime latency overhead}
\end{minipage}
\end{figure}

\textbf{\name{} cloud runtime overhead.} Fig. \ref{fig:cloud runtime latency overhead} shows the extra computation time added by our scheduler. As the offloading budget increases, scheduling overhead becomes more noticeable because each verification request has fewer uncached tokens and takes less time to compute, while the scheduling time stays almost constant. Even at the highest offloading budget, the overhead remains moderate around 25\%. For a typical working budget of 0.2, the overhead is negligible.

\begin{figure}[!t]
    \centering 
    \includegraphics[width=1.0\linewidth]{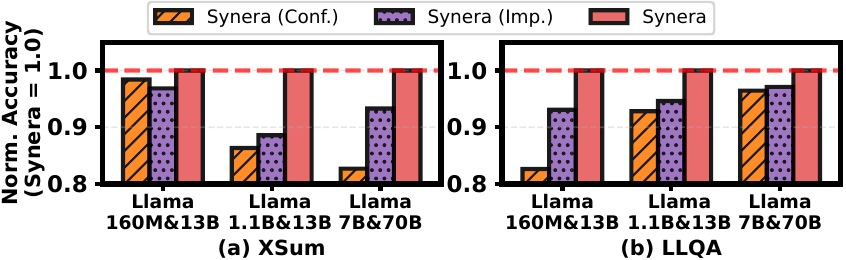} 
    \vspace{-20pt}
    \caption{Ablation Study of using only $P_{conf}$ and $P_{imp}$}\label{table: ablation study_conf_imp}
\end{figure}

\begin{table}[!t]
\centering
\resizebox{\linewidth}{!}{
\begin{tabular}{l|ccc}
\toprule
\textbf{Method} & \textbf{\makecell{Scheduling Latency \\per Token}} &  \textbf{\makecell{Energy \\per token}} & \textbf{\makecell{Compared to\\Edge-centric}} \\
\midrule
Edge-centric           & N/A       & 1.86 J & /   \\
Edge-centric (w/ EE)                    & N/A       & 1.62 J & -0.24 J  \\
\name{} (w/o EE)                & $<$0.5 ms    &  1.90 J & +0.04 J  \\
\name{} (w/o PI)                        & $<$0.5 ms    &  1.65 J & -0.21 J   \\

\name{}                         & $<$0.5 ms    &  1.83 J & -0.03 J   \\
\bottomrule
\end{tabular}
}
\caption{\name{} device runtime overheads.}\label{table: overheads_device}
\vspace{-12pt}
\end{table}

\subsection{\name{} with complementary methods}\label{evaluation: complementary method}
Table~\ref{table:complementary method} presents the speedup and accuracy of \name{} when combined with complementary SLM acceleration methods, such as bitsandbytes 4-bit and AWQ~\cite{lin2024awq} quantization. \name{} consistently delivers speedup and significant accuracy improvements ($\sim$40\%). These results demonstrate \name{}’s strong compatibility with existing SLM optimization.

\section{Related work}

\textbf{LLM on the Device/Cloud.}
Recent research has explored both on-device and cloud-based solutions for LLM inference. On-device approaches include model compression~\cite{li2023model, lin2024awq, frantar-sparsegpt}, filtering~\cite{yuan2022infi, li2020reducto}, table lookup~\cite{tang2023lut}, hardware-model co-design~\cite{lingneiwen1}, etc. These methods can enable computationally intensive LLMs running on resource‑constrained devices, albeit often at the cost of reduced accuracy. Meanwhile, cloud-based LLM systems are designed for high-throughput, low-latency inference, such as continuous batching~\cite{yu2022orca}, PagedAttention~\cite{vllm}, and chunked prefill~\cite{agrawal2024taming_sarathiserve}. 
In contrast, with respect to on‑device SLM, \name{} adopts an orthogonal strategy: it leverages cloud LLMs to improve the quality of on-device SLM while sustaining low latency. Synera is complementary with on-device approaches, such as model compression. At the same time, on the cloud side, Synera can readily harness the existing framework at the cloud runtime, thereby inheriting its benefits without additional complexity.

\begin{table}[!t]
\centering
\resizebox{\linewidth}{!}{
\begin{tabular}{l|ccc}
\toprule
\textbf{Method} & \textbf{\makecell{Speedup \\ (Norm.)}} & \textbf{\makecell{Rouge-1 Score \\ (Acc.)}} & \textbf{\makecell{Relative Accuracy \\ (Norm.)}} \\
\midrule
Edge-centric                & 1.0$\times$ & 13.81 & 1.0$\times$ \\
\name{}                     & 1.12$\times$ & 20.36 & 1.47$\times$ \\
\midrule
Edge-centric + BnB-4bit         & 1.0$\times$ & 12.89 & 1.0$\times$ \\
\name{} + BnB-4bit              & 1.18$\times$ & 18.84 & 1.46$\times$ \\
\midrule
Edge-centric + AWQ          & 1.0$\times$ & 10.90 & 1.0$\times$ \\
\name{} + AWQ               & 1.28$\times$ & 15.83 & 1.45$\times$ \\
\bottomrule
\end{tabular}
}
\caption{\name{} with complementary acceleration methods on XSum dataset across Yi-6B \& Yi-34B.}\label{table:complementary method}
\vspace{-10pt}
\end{table}

\textbf{Device–Cloud Synergy.}
Recent studies have demonstrated the benefits of device–cloud collaborative serving through data compression~\cite{deepcod}, multi‑stage processing~\cite{hung2018videoedge}, NN partitioning~\cite{gaowei_AgileNN,SPINN,chen2025collabtrans,li2019edge,huang2023re}, and small–large model synergy~\cite{edgefm,li2021appealnet,native_hybrid_renju_zixuhao,xiaozhulin_pasu}. However, most of these approaches are poorly suited to the unique challenges of generative LLMs. Instead, \name{} addresses LLM-specific challenges through SLM–LLM token‑level synergy, selectively offloading quality‑critical tokens to the cloud. Distinct from prior work, \name{} achieves advantageous and pragmatic integration with mature cloud frameworks such as vLLM. 

\textbf{Speculative Decoding.}
Speculative decoding~\cite{leviathan2023fast_google_speculative,chen2023accelerating_deepmind_speculative,miao2024specinfer,xu2025specee} accelerates LLM inference through a ``draft–verify'' paradigm: an SLM drafts candidate tokens that are then verified by an LLM. Building upon this, we further observe that token-level importance is not uniformly distributed, and only certain token chunks are quality‑critical. This insight reveals a new opportunity for device–cloud synergy: instead of offloading all verification steps, the system can selectively delegate these critical chunks to the cloud LLM, thereby opening new possibilities for efficient device–cloud synergy.

\section{Conclusion}

This paper presents \name{}, a device–cloud synergistic LLM serving system based on SLM–LLM synergy. \name{} identifies several previously overlooked dimensions of optimization and translates them into practical techniques: efficient offloading, stall‑free parallel inference, and scalable cloud batching. These innovations enable \name{} to deliver substantial gains in quality, cost-efficiency, and latency, offering a practical solution for synergistic LLM serving at scale.


\newpage
\bibliographystyle{ACM-Reference-Format}
\bibliography{reference}



\end{document}